\documentclass{aa}

   \usepackage{graphicx}

  \usepackage{txfonts}

\begin{document}

\title{European VLBI Network observations of fourteen GHz-Peaked-Spectrum radio sources at 5~GHz}
   \author{L. Xiang
          \inst{1}
           \and
           C. Reynolds\inst{2}
          \and
           R. G. Strom\inst{3}
           \and
           D. Dallacasa\inst{4,5}
 }

   \offprints{L. Xiang}

   \institute{National Astronomical Observatories/Urumqi Observatory, CAS,
   40-5 South Beijing Road, Urumqi 830011, PR~China\\
   \email{liux@ms.xjb.ac.cn}
 \and Joint Institute for VLBI in Europe, Postbus 2, 7990 AA Dwingeloo, The Netherlands\\
 \email{reynolds@jive.nl}
\and ASTRON, Postbus 2, 7990 AA Dwingeloo;
and Astronomical Institute, University of Amsterdam, The Netherlands\\
 \email{strom@astron.nl}
 \and Dipartimento di Astronomia, Universit\`a di Bologna, via Ranzani 1, I-40127 Bologna, Italy\\
             \email{daniele.dallacasa@unibo.it}
             \and Istituto di Radioastronomia -- INAF, via P. Gobetti 101, 40129 Bologna, Italy\\
}

 \date{Received 28 July 2005 / Accepted 30 December 2005}


  \abstract {We present the results of EVN polarization observations of
fourteen GHz-Peaked-Spectrum (GPS) radio sources at 5~GHz. These
sources were selected from bright GPS source samples and we
aimed at finding Compact Symmetric Objects (CSOs). We have
obtained full polarization 5~GHz VLBI observations of 14 sources
providing information on their source structure and spectral
indices. The results show that two core-jet sources 1433$-$040 and
DA193, out of 14 GPS sources, exhibit integrated fractional
polarizations of 3.6\% and 1.0\% respectively. The other 12
sources have no clear detection of pc-scale polarization. The
results confirm that the GPS sources generally have very low
polarization at 5~GHz. The sources 1133+432, 1824+271 and
2121$-$014 are confirmed as CSOs. Three new CSOs 0914+114,
1518+046 and 2322$-$040 (tentative) have been classified on the
basis of 5~GHz images and spectral indices. The sources 1333+589,
1751+278 and 2323+790 can be classified either as compact doubles, and
then they are likely CSO candidates or core-jet sources; further
observations are needed for an appropriate classification; 0554$-$026,
1433$-$040 and 1509+054 are core-jet sources. In addition, we estimate
that a component in the jet of quasar DA193 has superluminal motion of
$3.3\pm0.6\, h^{-1}c$ in 5.5 years.

}

   \titlerunning{EVN observations of fourteen GHz-Peaked-Spectrum (GPS) radio
sources at 5~GHz}
\authorrunning{L. Xiang et al.}

\maketitle

\section{Introduction}

Compact Symmetric Objects (CSOs) are of particular interest in the
study of the physics and evolution of active galaxies as
introduced in our previous papers (Xiang et al. 2002, 2005,
hereafter paper I, II). A number of CSOs have been found to be
very young radio sources with ages of several centuries (Fanti
2000; Polatidis \& Conway 2003; Gugliucci et al. 2005). How and
why the radio emission has been recently triggered is still a matter
of debate. There is some evidence that the host galaxies of CSO
radio sources may be in a final (or even post-) merger stage, but
no sound conclusion can be drawn given the small number of known
CSOs. A larger sample of CSOs and more detailed studies of their
radio properties, to be compared with those of old and extended
radio galaxies, are fundamental steps for a better understanding
of this phenomenon.

Models have been proposed to explain the evolution of individual
small and young radio galaxies: their radio emission grows in a
self-similar way whereby the small CSOs become the extended
radio galaxies about 3 orders of magnitude larger and about 4
orders of magnitude older than CSOs (e.g. Fanti et al. 1995;
Readhead et al. 1996; Snellen et al. 2000). Support for this
hypothesis has been found by Owsianik \& Conway (1998) who
reported on the detection of outward motion of the hot-spots of a
few CSOs, and by Murgia et al. (1999) who found that the radiative
ages of small sources are consistent with a young age.

CSOs have their radio axes roughly on the plane of the sky and are
generally associated with elliptical galaxies. As a consequence of
Doppler dimming, their jets are generally very rarely observed in
this case, and the morphology is characterized by two bright
lobes, possibly hosting hot-spots. Given their small size either
Synchrotron Self-Absorption (SSA) or Free-Free absorption (FFA) or
both take place, producing the typical peaked radio spectrum
around 0.3-10~GHz in the observer's frame. GHz-Peaked Spectrum
(GPS) sources are also objects where HI absorption at the host
redshift has been frequently found (e.g. Vermeulen et al. 2003).
In the case where the radio axis is aligned to the line of sight,
the typical radio structure is a one-sided core-jet with Doppler
boosting responsible for the amplification of the approaching side
of the radio source.

Compact two-sided sources have been found in samples based on the
spectral shape: CSOs are in the Pearson \& Readhead (PR, 1988) and
in the Caltech-Jodrell Bank (e.g. CJ-1; Xu et al. 1995) surveys of
flat spectrum sources where the spectral peak is between the two
frequencies used to define the source spectrum. Similar work has
been done by Taylor \& Peck (2003). In a flux-limited complete
sample such as PR+CJ1 (200 sources with $S_{5GHz}>$ 0.7 Jy and
$\delta>35^{\circ}$) 14 sources have been classified as CSOs.
Among them, 10 also fulfill the definition of GPS sources. It is
therefore efficient to search for CSOs in GPS samples (see paper
I, II). GPS sources share the same properties as CSOs: they are
generally smaller than 1~kpc and are young radio sources (Murgia
2003).

GPS radio sources make up a significant fraction of the bright
radio source sample ($\approx$ 10\%) but they are not well
understood. The GPS sources are powerful
($P_{\rm1.4~GHz}\geq\rm10^{25}W\,Hz^{-1}$), compact ($\leq
1$~kpc), and have convex radio spectra. Only about 18\%
(Stanghellini et al. 2005) of GPS sources show extended radio
emission ($> 1$~kpc), and it is diffuse and very faint, possibly
the relic of an earlier radio activity. Most GPS sources appear to
be truly compact and isolated (O'Dea 1998).

GPS radio galaxies show very low polarization (about 0.2\% at
5~GHz, O'Dea 1998). The low integrated polarization may originate
in a medium with a large Faraday depth located around the radio
source. While the vast majority of GPS galaxies are found
to be unpolarized at cm wavelengths, GPS quasars show relatively higher
fractional polarization (Dallacasa 2004) consistent with
(possibly slightly lower than) flat spectrum radio quasars. This
implies that the galaxies are depolarized more than the quasars as
is also expected (Cotton et al. 2003) in terms of Unified Scheme
models. Higher frequency observations are required to determine
the differences between GPS galaxies and quasars in polarization.
The very low level of polarization often makes it difficult to
measure reliable Rotation Measures (RMs) in GPS galaxies. The few
sources with such measurements possess large RM values.

``Bright'' GPS source samples (de Vries et al. 1997, Stanghellini
et al. 1998) comprise a few tens of objects, and a couple of them
either have never been imaged with VLBI or posses low dynamic
range images. From the aforementioned samples, we have observed 22
sources in a number of VLBI projects carried out with the EVN
(European VLBI Network): paper I reports on the first observing
run (1999 November 13 at 1.6~GHz and 2000 May 29 at 2.3/8.4~GHz)
while paper II reports on the second session (2002 July 4 at
2.3/8.4~GHz). From the multi-frequency VLBI images five CSOs and
nine CSO candidates have been found.

In these follow up observations at 5~GHz, we aim at higher dynamic
range imaging of the GPS sources, to confirm the CSO candidates
found in earlier papers, to find new CSOs from the foremetioned
bright GPS samples, and, possibly, to measure the linear
polarization. Our search for CSOs from the GPS samples is
complementary to that by the COINS group (Taylor \& Peck 2003):
there the search for CSOs is based on samples consisting mainly of
flat spectrum radio sources, although also a GPS may appear as a
flat or even inverted spectrum object if the peak flux density
occurs close to the highest frequency of the selection.

\section{Observations and data reduction}

The observations were carried out on 30 October 2004 at 5~GHz
using the MK5 recording system with a bandwidth of 16 MHz and
sample rate of 128 Mbps in dual circular polarization. The EVN
antennae in this experiment were Effelsberg, Westerbork (phased
array), Jodrell, Medicina, Noto, Onsala, Torun, Hartebeesthoek,
Urumqi and Shanghai. All stations contributed useful fringes.
Snapshot observations of 14 sources (see Table~\ref{table:gps}) in
a total of 24 hours were done. The calibration sources OQ208 and
DA193 were observed with three 6-minute scans each. Most target
sources were observed in more than ten 6-minute scans. The data
correlation was completed at JIVE in March 2005.

Data from the Westerbork Synthesis Radio Telescope (WSRT) taken in
parallel with the EVN session were processed to determine
integrated properties for all the sources (including the source
3C48). The data were reduced in a standard way using the local
NEWSTAR reduction package. All the target sources were found
pointlike at the WSRT resolution ($\sim$5$\arcsec$) and their
total flux densities are reported in Table~\ref{table:wb}.

The total flux densities of some sources were also measured shortly
after the VLBI experiment in Effelsberg (5 sources; 3C286 was
observed with 7.4 Jy). These values are listed in
Table~\ref{table:gps} as well.

Examination of the WSRT visibilities demonstrated that the array had
been phased up well, so no further calibration was required to
process the data. The calibrator OQ208 (believed to be
unpolarized) was used to determine the instrumental polarization
characteristics of each element in the array (dipole setting error
and ellipticity). For each source, the Stokes parameters $I$, $Q$,
$U$ and $V$ were determined from maps. The values for Stokes $I$
have been scaled to the flux density of 3C48 (taken to be 5.48
Jy). Stokes $Q$, $U$ and $I$ were used to determine the degree of
linear polarization, $m$ ($=(Q^2+U^2)^{0.5}/I$), and position
angle of the electric vector (EVPA), $\chi$
($=\frac{1}{2}\arctan(U/Q)$). The degree of circular polarization,
$m_V$($=V/I$), was also measured.

For most of the target sources observed, the degree of polarization
was found to be small, as has been seen in GPS sources in the past
(O'Dea 1998). Examination of the $m$ values for all sources showed
that the majority had a degree of polarization near 0.6\%, and with
similar values of $\chi$. It seems most likely that the majority
of weakly-polarized sources would cluster around $m=0$, so it was
concluded that the polarization of OQ208 in our measurement is not
exactly zero. The polarization data were thus recalibrated using
the average $m$ (0.6\%) and $\chi$ ($85^\circ$) found above, and
these were taken to be definitive. The values for 15 sources can
be found in Table~\ref{table:wb}. The probable error in $m$, based
upon internal consistency and experience with the WSRT as a
polarimeter, is estimated to be $\pm0.4$\%; the error in $\chi$
(Table~\ref{table:wb}) depends upon this and the measured value of
$m$ (no value of $\chi$ is given where the error is twice $m$ or
more). Practically all of the circular polarization ($m_V$) values
measured are consistent with there being no significant circular
polarization in these sources. The value we find for OQ208 is
consistent with the upper limit established by Homan, Attridge and
Wardle (2001).

\subsection{The VLBI data}
The Astronomical Image Processing System (AIPS) has been used for
editing, {\it a priori} calibration, fringe-fitting,
self-calibration and imaging of the VLBI data. The polarization
calibration was carried out following the standard procedure in
AIPS. The instrumental polarization was removed by LPCAL run on
\object{OQ208} which was assumed to be unpolarized and compact on
a subset of baselines. The polarization angles on the VLBI images
were calibrated by comparing the \object{DA193} polarization angle
in the integrated VLBI data with that of the Westerbork
interferometric data, taken simultaneously to the VLBI
observation. The correction was subsequently checked by comparing
the WSRT and VLBI observations of the target source
\object{1433$-$040} which gave a results agreeing within a few
degrees. The Westerbork data were in turn checked using
observations of 3C\,48 which was measured to have an EVPA of
$115^{\circ}$, in reasonable agreement with an Effelsberg
single-dish result of 104$^{\circ}$ for 3C\,48 observed at
4.85~GHz on 7 October 2004 (W. Reich, private commun.).

\section{Results and comments on individual sources}

Results from the 5~GHz VLBI observations are presented in this
section. Basic information and some total flux densities of the
sources appear in Table~\ref{table:gps}, the integrated properties
of the sources from the WSRT data appear in Table~\ref{table:wb},
the parameters derived from the VLBI images are summarized in
Table~\ref{table:ex004}. We comment on the results for each source
and then give a short discussion. We use $S \propto \nu^{-\alpha}$
in this paper to define the spectral index.

   \begin{table*}

         \caption[]{GPS sources. Columns 1 through 13 provide source name, optical identification (G: galaxy,
Q: quasar, EF: empty field), optical magnitude (`V',`R' are the
bands in the optical, `r' band--defined by Stanghellini et al.
1993), redshift (those with * are a photometric estimate by
Heckman et al. 1994), linear scale factor pc/mas [$H_{0}=100\, h\,
\rm km\,s^{-1}\, Mpc^{-1}$ and $q_{0}=0.5$ have been assumed],
maximum VLBI angular size from paper I, II or this observation,
maximum VLBI linear size, 5~GHz total flux density measured at
Effelsberg, low frequency spectral index, high frequency spectral
index, turnover frequency, peak flux density and references for
the spectral information (ref: 1, de Vries et al. 1997; 2,
Stanghellini et al. 1998; 3, Dallacasa et al. 2000; 4, O'Dea et
al. 1996; 5, Spoelstra et al. 1985).}

\begin{center}

           \begin{tabular}{ccccccccccccc}

            \hline

            \hline

            \noalign{\smallskip}

            1&2 &3 &4 & 5& 6 & 7 & 8&9 &10 & 11 & 12 & 13\\

source & $id$ & $mag$ & $z$ & $pc/mas$ & $\theta$ & $L$ & $S_{\rm{Ef}}$  & $\alpha_{l}$ & $\alpha_{h}$ & $\nu_{m}$ & $S_{m}$ & ref \\

& & & & & [mas] & [pc] & [Jy]& & & [GHz] & [Jy] & \\

            \noalign{\smallskip}

            \hline

            \noalign{\smallskip}

\object{0554$-$026} & G   & 18.5V  & 0.235 & $2.37h^{-1}$ & 30  & $71h^{-1}$ &     & -1.07 & 0.63  & 1.0 & 0.8 & 1 \\

\object{0914+114} & G   & 20.0r  & 0.178 & $1.95h^{-1}$ & 150 & $293h^{-1}$&0.11 & -0.1  & 1.6   & 0.3 & 2.3 & 2 \\

\object{1133+432} & EF  &        &       &             & 40  &            &0.45  & -0.60 & 0.6   & 1.0 & 1.4 & 1 \\

\object{1333+589} & EF  &        &       &             & 20  &            &0.67 & -0.84 & 0.52 & 4.9 & 0.73& 3 \\

\object{1433$-$040} & G  & 22.3r  &       &             & 20  &            &      &     &     &    &   & 4,5\\

\object{1509+054} & G   & 16.2V  &  0.084 & $1.06h^{-1}$  & 17  & $18h^{-1}$ &    & -1.72 & 0.46  & 11 & 0.77& 3 \\

\object{1518+046} & Q   & 22.2R  & 1.296 & $4.3h^{-1}$ & 160 &$688h^{-1}$ &1.06  & -0.6  & 1.3  & 0.9 &4.58 &2  \\

\object{1751+278} & G   & 21.7R  & 0.86* & $4.2h^{-1}$ & 50  &$210h^{-1}$ &      & -0.27 & 0.57  & 1.4 &0.6 &1  \\

\object{1824+271} & G?  & 22.9R  &       &             & 45  &            &      & -0.39 & 0.75  & 1.0 & 0.4 & 1 \\

\object{2121$-$014} & G   & 23.3R  & 1.158 & $4.3h^{-1}$ & 88  &$378h^{-1}$ &      & -0.56 & 0.75  & 0.5 & 1.8 & 1 \\

\object{2322$-$040} & G  & 23.5R  &       &             & 65  &            &     & -0.42 & 0.75  & 1.4 & 1.3 & 1 \\

\object{2323+790} & G   & 19.5V  &       &             & 32  &            &0.43  & -0.3  &0.75   &1.4  &1.2 & 1 \\

\object{DA193}    & Q   & 18.0V  & 2.365 & $4.0h^{-1}$  & 4   & $16h^{-1}$ &      & -1.6  & 0.6  & 9.5 & 7.8  & 2  \\

\object{OQ208}    & G   & 14.6r  & 0.077 & $1.0h^{-1}$ & 10  & $10h^{-1}$ &      & -1.5  & 1.6  & 4.9 &2.76  & 2 \\

\noalign{\smallskip}

            \hline

          \end{tabular}

          \end{center}

          \label{table:gps}

   \end{table*}

   \begin{table*}

         \caption[]{Integrated properties of the sources from the WSRT
          data, values of $\chi$ are likely to be meaningless for objects with
          $m \leq 0.4\%$.}

\begin{center}

          \begin{tabular}{ccccc}

            \hline

            \hline

            \noalign{\smallskip}

             1&2 &3 &4 & 5\\

source & $I$ & $m$ & $\chi\pm\sigma_\chi$ & $m_V$  \\

& [mJy] & [\%] & [$^\circ$] & [\%]  \\

            \noalign{\smallskip}

            \hline

            \noalign{\smallskip}

0554$-$026 & $171\pm6$ & 0.3 & $18\pm43$  & 0.1 \\

0914+114 & $118\pm9$ & 0.4 & $94\pm33$  & 0.1  \\

1133+432 & $480\pm15$ &  0.2 & - & -0.0   \\

1333+589 & $720\pm42$ &  0.4 &  $141\pm30$ & -0.0   \\

1433$-$040 & $246\pm15$ & 3.8 & $134\pm5$ & -0.3  \\

1509+054 & $688\pm63$ & 0.7 & $173\pm17$  & 0.1  \\

1518+046 & $994\pm52$ & 0.7 & $176\pm17$  & 0.0 \\

1751+278 & $260\pm12$ & 0.3 & $164\pm41$ & -0.1  \\

1824+271 & $111\pm5$ & 0.3 & - & -0.0  \\

2121$-$014 & $327\pm13$ & 0.4 &  $29\pm30$ &  0.0  \\

2322$-$040 & $548\pm15$ & 0.1 & - & -0.1  \\

2323+790 & $438\pm27$ & 0.5 &  $93\pm25$ & -0.0  \\

3C48  & $5480\pm50$ & 3.1 & $115\pm5$ & -0.0  \\

DA193  & $5232\pm64$ & 1.3 & $149\pm9$ &  0.1 \\

OQ208  & $2609\pm15$ & 0.6 & $175\pm19$ & 0.0  \\

\noalign{\smallskip}

            \hline

          \end{tabular}

          \end{center}

         \label{table:wb}

   \end{table*}

   \begin{table*}

         \caption[]{Parameters of the components in the VLBI images at 5~GHz. The columns give: (1)
source name and possible classification (CSOc: Compact Symmetric
Object candidate, cj: core-jet); (2) total image intensity
measured with IMEAN at 5~GHz; (3) component identification labled
to Xiang et al. 2002, 2005; (4),(5) peak (per beam) and integral
intensity of component at 5~GHz measured with JMFIT; (6),(7)
major/minor axes (with error $\leq 10\%$) and position angle of
component at 5~GHz measured with JMFIT; (8),(9) distance and
position angle relative to the first component; (10) brightness
temperature of component; (11) integrated fractional polarization
in source.}

\begin{center}

        \begin{tabular}{ccccccccccc}

          \hline

           \hline

            \noalign{\smallskip}

             1&2 &3 &4 & 5& 6 & 7 & 8&9&10&11 \\

Name    & $S_{vlbi}$ & & $S_p$    & $S_{int}$ & $\theta_{1}\times\theta_{2}$ & $PA$  & $d$ && $T$& $m$  \\

class   & [mJy]      &        & [mJy]    & [mJy]          & [mas]    & [$^{\circ}$] & [mas]  &[$^{\circ}$]&

[$10^{7}K^{\circ}$]  & [\%]  \\

            \noalign{\smallskip}

            \hline

            \noalign{\smallskip}

\object{0554$-$026} & $175\pm3$ & AB  & $50.8\pm0.7$ & $114.9\pm2.1$& 7.1$\times$3.0& $0.1\pm0.5$ &    0       &    & 2.0 &$\leq0.4$ \\

cj         &           & C   & $17.6\pm0.6$ & $77.3\pm3.4$ & 10.7$\times$3.9&$70.4\pm1.3$& $0.4\pm0.1$ &$-69.7\pm4.7$& 0.6&      \\

\object{0914+114} & $116\pm2$ & A   & $27.2\pm0.8$ & $43.0\pm2.0$ & 6.5$\times$3.3 &$0.6\pm1.8$  & 0   &  &  0.6&   \\

CSO      &             & B &               &$\leq4.3\pm1.9$  &   &   &    &  &&     \\

         &             & tail &            &$20\pm1.3$       &   &   &    &  &&     \\

         &             & C   & $25.2\pm0.8$ & $46.1\pm2.2$ & 6.7$\times$3.7 &$4.1\pm2.1$  & $84.3\pm0.1$ &$81.4\pm0.1$ &0.6 &     \\

         &             & D   &            & $\leq3.8\pm1.5$  &   &   &    &  &&     \\

\object{1133+432} & $416\pm11$   & A   &           & $290\pm14$ &                  &      &    & &&    \\

CSO      &            & B1  & $44.5\pm0.4$ &$67.4\pm0.9$& 3.2$\times$2.1& $35.3\pm0.8$& 0 && &       \\

         &            & B2  & $25.5\pm0.4$ &$53.3\pm1.1$& 3.3$\times$2.8& $106.8\pm3.3$& $2.7\pm0.1$&$1.3\pm0.5$& &       \\

\object{1333+589} & $580\pm55$   &N  &$288.5\pm2.4$ &$331.6\pm4.5$& 3.4$\times$3.2& $45.1\pm4.8$& 0 &&  &        \\

 CSOc/cj &            & S  &$182.0\pm2.3$ &$225.0\pm4.7$& 3.7$\times$3.2& $125.7\pm4.0$ & $12.7\pm0.1$ &$197.5\pm0.5$&&    \\

\object{1433$-$040}&$206\pm10$   &core &$66.5\pm0.3$ & $125.8\pm0.8$& 6.1$\times$2.8& $177.8\pm0.2$& 0 &&   & 3.6     \\

 cj      &            & jet &$14.1\pm0.3$ & $67.7\pm1.5$ & 9.6$\times$4.5& $17.2\pm0.9$ & $1.4\pm0.2$&$12.4\pm3.5$& & 3.6     \\

\object{1509+054} &        &  E & $76.9\pm1.5$ & $108.3\pm3.3$ & 4.6$\times$3.7& $2.3\pm3.5$  & 0  &&2.0 &     \\

 cj      &        &  W & $389.0\pm1.5$& $548.5\pm3.3$& 4.9$\times$3.5& $1.5\pm0.5$  & $5.2\pm0.1$&$-89.7\pm0.4$&10.2&    \\

         & $682\pm31$   &  A & $80.8\pm1.5$ & $91.8\pm2.7$ & 3.8$\times$1.2& $7.1\pm0.5$  & 0  &&6.5 &     \\

         &            &  B & $317.3\pm1.4$& $464.0\pm3.2$& 4.2$\times$1.4& $7.4\pm0.1$  & $5.3\pm0.1$&$-89.2\pm0.1$&25.2& $\leq0.4$      \\

         &            &  C & $38.0\pm1.4$ & $92.8\pm4.5$ & 5.2$\times$2.0& $13.6\pm1.3$ & $7.6\pm0.1$&$-81.6\pm0.1$&2.9&      \\

         &            &  D & $27.3\pm1.4$ & $33.2\pm2.8$ & 3.9$\times$1.3& $7.6\pm1.6$  & $10.4\pm0.1$&$-97.1\pm0.1$&2.1&     \\

\object{1518+046} & $869\pm9$    &  A & $135.4\pm1.6$& $268.6\pm4.4$& 6.7$\times$3.9& $1.9\pm0.9$  &  0 &&  7.0 & $\leq0.5$         \\

CSO/MSO  &            &  B & $49.7\pm1.6$ & $108.9\pm4.8$& 6.7$\times$4.3& $4.3\pm2.9$  &  $10.4\pm0.1$&$39.9\pm0.1$&2.6&\\

         &           &  C  & $205.6\pm1.7$& $249.8\pm3.3$& 5.2$\times$3.1&$178.9\pm0.6$& $135.8\pm0.1$&$206.7\pm0.1$ &10.5&  $\leq0.5$        \\

         &           &  D  & $97.0\pm1.6$ & $238.0\pm5.2$& 8.0$\times$4.1& $23.1\pm0.9$& $133.2\pm0.1$ & $207.0\pm0.1$&4.9&\\

\object{1751+278} & $270\pm10$  &  A  & $163.4\pm0.5$& $243.4\pm1.1$& 5.3$\times$3.5& $175.4\pm0.3$& 0  && 7.2 & $\leq0.4$         \\

 CSOc/cj &           &  B  & $11.7\pm0.8$ & $20.6\pm2.2$ & 5.8$\times$3.7& $7.7\pm6.6$ & $23.3\pm0.2$&$227.5\pm0.1$&0.5 &         \\

\object{1824+271} & $115\pm6$ & A & $48.5\pm0.2$ & $74.3\pm0.5$ & 6.3$\times$3.9& $177.3\pm0.4$& 0  &&  &    \\

 CSO     &            &  B & $19.7\pm0.2$ & $31.4\pm0.5$ & 6.1$\times$4.2& $175.6\pm1.1$& $21.6\pm0.1$&$-82.6\pm0.1$& &      \\

\object{2121$-$014} & $317\pm9$  & A1 & $99.4\pm0.8$& $154.2\pm1.8$& 6.8$\times$3.6& $2.3\pm0.5$ & 0 &&  4.0 &     \\

 CSO     &            & A2 & $8.3\pm0.7$  & $42.0\pm4.3$ & 9.8$\times$8.3 &$33.0\pm21.4$&$2.8\pm0.7$&$57.4\pm2.1$& 0.3 &       \\

         &            & B  & $7.4\pm0.8$  & $13.3\pm2.2$ & 6.6$\times$4.3 & $0.6\pm11$ &$16.8\pm0.6$&$86.5\pm1.8$   & 0.3 &     \\

         &            & C1 & $33.0\pm0.9$ & $33.7\pm1.6$ & 4.5$\times$3.6 & $2.9\pm4.6$&$60.8\pm0.5$&$82.0\pm0.4$&13.4 &     \\

         &            & C2 & $33.4\pm0.8$ & $73.7\pm2.6$ & 9.2$\times$3.8 & $1.8\pm1.0$&$61.4\pm0.5$ &$84.4\pm0.4$&13.5&     \\

\object{2322$-$040}& $560\pm15$  & A  & $290.2\pm1.5$& $429.1\pm3.4$& 5.1$\times$2.7 &$175.0\pm0.3$ & 0 &&   &         \\

 CSO     &           & tail&$24.4\pm1.5$ & $50.6\pm4.2$ & 7.2$\times$2.6 &$3.7\pm2.1$  & $6.7\pm0.2$&$192.5\pm0.2$& &          \\

         &            &  B & $40.7\pm1.5$ & $78.3\pm4.0$ & 6.1$\times$2.8 &$157.8\pm1.7$& $37.3\pm0.2$&$176.4\pm0.2$&&    \\

\object{2323+790} & $409\pm13$  &  A1 & $73.6\pm0.3$ & $229.4\pm1.3$& 9.9$\times$3.7 &$10.8\pm0.2$  & 0  &&  &         \\

 CSOc/cj &           &  A2 & $110.0\pm0.3$& $123.0\pm0.6$& 4.1$\times$3.3 &$172.6\pm0.5$ & $0.2\pm0.1$&$143.4\pm1.2$&&          \\

         &           &  B  & $30.3\pm0.3$ & $45.5\pm0.7$ & 4.8$\times$3.7 &$25.0\pm1.7$  & $17.1\pm0.1$ & $-76.8\pm0.1$&&   \\

         &            &  C & $8.7\pm0.3$  & $24.0\pm1.1$ & 7.8$\times$4.2 &$9.3\pm2.3$   & $22.5\pm0.2$ & $-77.5\pm0.3$&&   \\

\object{DA193}    &$5272\pm172$ & core & $3415\pm7$&$3943\pm13$&3.8$\times$1.5& $22.7\pm0.1$  & 0 &&  686.5 & 1.0       \\

 cj      &           & jet  & $951\pm7$ &$1300\pm14$&3.8$\times$1.8& $21.2\pm0.3$  & $2.1\pm0.1$&$-48.3\pm0.1$& 191.0& 1.0   \\

\object{OQ208}    &$2656\pm96$ & NE & $1728\pm7$& $2342\pm16$ & 4.7$\times$2.8& $3.3\pm0.3$  & 0 &&  56.5 &       \\

 CSO     &          & SW & $173\pm7$ & $236\pm16$  & 4.0$\times$3.3& $17.1\pm9.3$ & $6.5\pm0.2$ &$230.2\pm0.2$& 5.7&     \\

\noalign{\smallskip}

            \hline

        \end{tabular}

        \end{center}

         \label{table:ex004}

   \end{table*}

\subsection{\object {0552+398} (\object {DA193}, \object {NVSS J055530+394848}, \object {WMAP 100})}

\begin{figure}
      \includegraphics[width=7cm]{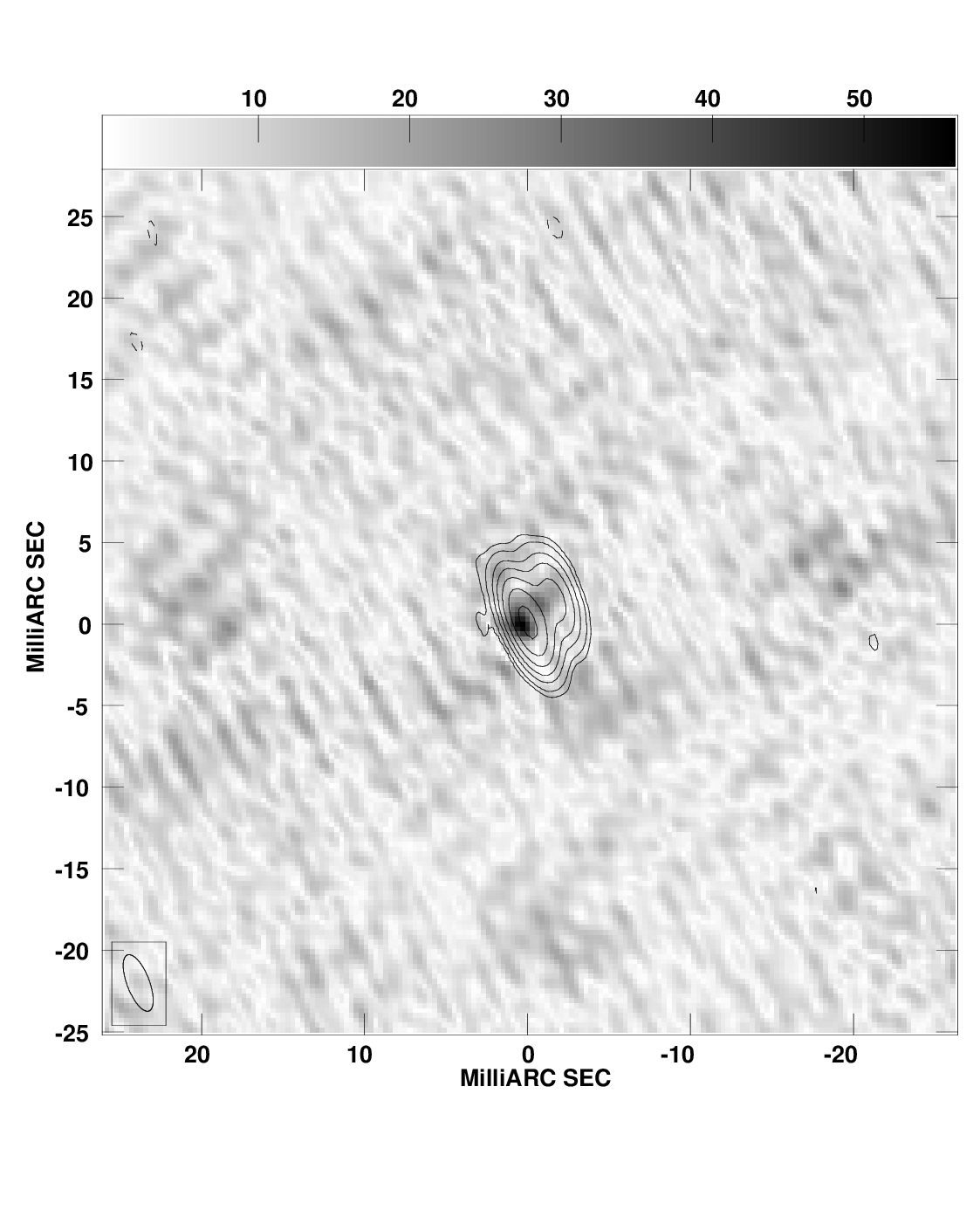}
      \caption{DA193 at 5~GHz, the restoring beam is $3.7\times1.4$
        mas in PA $21.0^{\circ}$, the contours are total intensity with 34 mJy/beam times -2, -1, 1, 2, 5, 10, 20, 40, 80,
        and the peak of 3433 mJy/beam.
        The overlay in grey is linearly polarized intensity, with the peak of 56 mJy/beam.}
      \label{da193_ip}
   \end{figure}

The low-polarization quasar DA193 is a non-blazar source
associated with an X-ray source \object {1WGA J0555+3948}, and has
been classified as a GPS source (Stanghellini et al. 1998) with a
spectral turnover at 9.5~GHz. This radio loud quasar is extremely
compact, is only resolved with VLBI, and is often used as a VLBI
calibrator. The total intensity ($I$) map at 5~GHz
(Fig.~\ref{da193_ip}) clearly shows a core and a jet. The core is
very compact with about 70\% of the total flux density, the jet
component is also compact with about 30\% of the total flux
density. The jet position angle is $-48^{\circ}$ in the north-west
direction, roughly consistent with $-53^{\circ}$ in a VSOP map at
5~GHz (Scott et al. 2004). The distance between the jet component
and the core is 2.1 mas in our 5~GHz VLBI observation, which is
greater than the 0.7 mas in the previous VSOP image. If the
component detected here is the same as that detected in the VSOP
image, then the increase over about 5.5 years corresponds to an
apparent velocity of $3.3\pm0.6\, h^{-1}c$. It has previously been
suggested that the jet has a superluminal motion of about
$2.2\pm0.7\, h^{-1}c$ (Lister \& Marscher 1998). The source is
known to have relatively smooth flux density variability; the
5~GHz flux density has varied from 6.31 Jy (Stanghellini et al. 1998)
to 5.23 Jy in this observation. At the epoch of the VSOP result
(observed in March 1999) the core and the jet flux densities are
4.72 Jy and 2.57 Jy respectively, giving a total of 7.29 Jy if
they were not over-estimated. So the jet has advanced with
decreasing flux and keeping almost the same PA from March 1999 to
October 2004. The VLA/VLBA monitoring data from  1999 to 2005 are
decreasing for DA193 from 6.3 Jy to 5.1 Jy at 5~GHz.

The polarization VLBI image shows a weak feature, with integrated
fractional polarization of 1\% and $\chi= 149^\circ\pm9^\circ$
($-31^\circ\pm9^\circ$). The fractional polarization measured at
the WSRT is 1.3\%. If we consider the integrated values measured
at the VLA at various frequencies ranging from 5.0 to 43 GHz
(www.vla.nrao.edu/astro/calib/polar/), it is clear that the
fractional polarization is generally of the order of 1\% or
slightly higher with a substantial amount of Faraday Rotation,
although it is difficult to estimate.

\subsection{\object {PKS 0554$-$026} (\object {NVSS J055652$-$024105})}

The 5~GHz VLBI image shows a compact core, with a jet or diffuse
emission within 20~pc (Fig.~\ref{0554-026_i}). The total flux
density measured at the WSRT has been fully accounted for in this
image. The total flux density has decreased from 290 mJy (observed
1980, Wright \& Otrupcek, 1990; Parkes Catalogue) to 171 mJy in
this observation, indicating the source is variable and that the
structure has to be interpreted in terms of a core-jet. The
structure at 5~GHz is similar to  that found at 2.3/8.4~GHz (paper
II) in general. There is no clear detection of polarized flux.

\subsection{\object {0914+114} (\object {PKS 0914+11}, \object {NVSS J091716+111336})}

\begin{figure}
      \includegraphics[width=7cm]{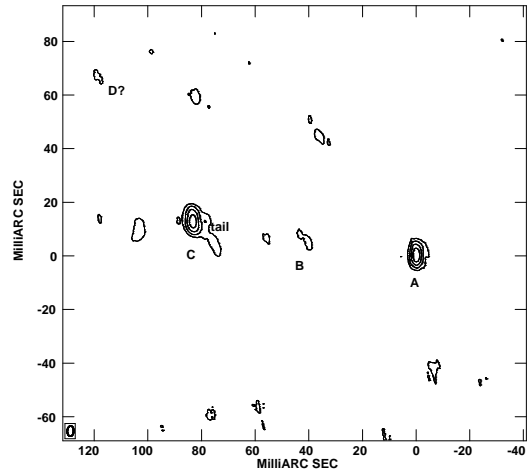}
      \caption{0914+114 at 5~GHz, the restoring beam is $4.7\times2.9$
        mas in PA $-1.5^{\circ}$,  the contours are 1.6 mJy/beam times -2, 1, 2, 4, 8, with the peak of 28 mJy/beam.
        }
      \label{0914+114_i}
   \end{figure}

\begin{figure}
      \includegraphics[width=7cm]{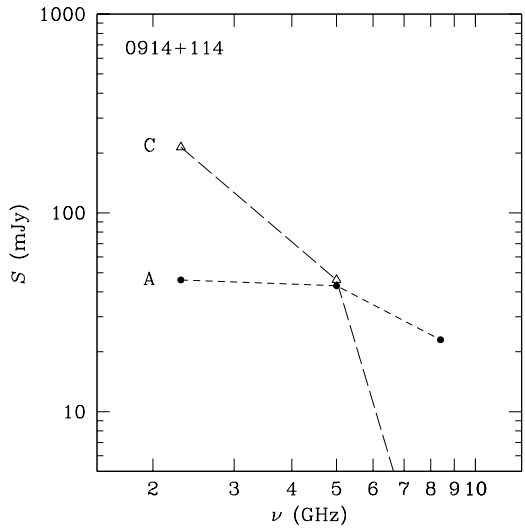}
      \caption{Component spectra of source 0914+114 from the VLBI data at 2.3, 5.0, and 8.4~GHz, where flux
      of component C at 8.4~GHz is $\leq 1$ ~mJy.}
      \label{0914+114_spec}
   \end{figure}

The 5~GHz VLBI image (Fig.~\ref{0914+114_i}) exhibits a symmetric
double source, with hints of a ``tail'' associated with the
eastern lobe. From the relative separation between two main components
detected in the 5~GHz VLBI image, it is clear that they correspond
to components `A' and `C' in paper II. An upper limit of
$4.3\pm1.9$ mJy may be associated at the position of component `B'
while it seems there is no likely detection at the position of
component `D' (paper II). Here we tentatively register the single
component at 8.4~GHz (it was assumed to be component `C' in paper
II) as component `A' in the 5~GHz image by considerations based on
its spectrum. Such an identification means that the spectral
indices of both components `A' and `C' between 5 and 8.4~GHz are
steeper than between 2.3 and 5~GHz
(Fig.~\ref{0914+114_spec}), which is typical of the components in
CSOs above the turnover frequency. In this case component `A' has
a flat spectrum ($\alpha=0.09$) between 2.3 and 5~GHz and this
allows its detection at 8.4~GHz, while component `C' would
have a rather steep spectrum between 2.3 and 5~GHz and then it
could reasonably be expected to be undetected at 8.4~GHz. The
possibility that the single component detected at 8.4~GHz has to
be associated with component `B' in paper II, is ruled out by the
upper limit of 4.3 mJy found in the present data. Because component `B'
has a flat spectrum of $\alpha=0.23$ between 2.3 and 8.4~GHz.

From the inverted spectra of the components and their symmetrical
structure, we classify the source as a CSO. The total flux density
has decreased from 140 mJy (observed 1979, Parkes Catalogue 1990)
to 118 mJy in this observation. For our observations from
2.3/8.4~GHz to 5~GHz separated by 2.4 years, source variability
may have influenced the spectral index estimates, but the error
introduced by this should be small comparing to the long term
variation of $\sim$16\%. The total flux density at 5~GHz is
recovered in the VLBI image. No polarization is detected in the
5~GHz VLBI observation.

\subsection{\object {1133+432} (\object {NVSS J113555+425844})}

\begin{figure}
      \includegraphics[width=7cm]{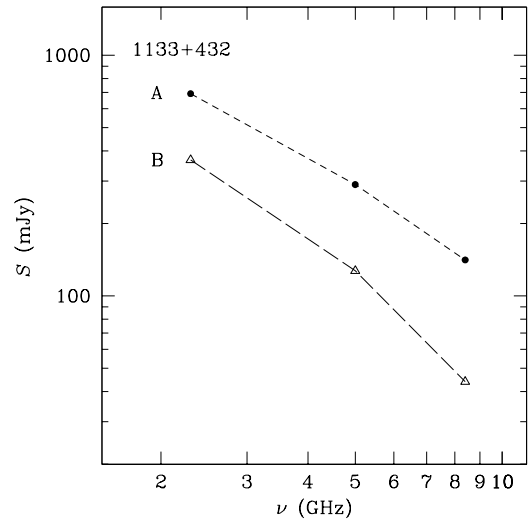}
      \caption{Component spectra of source 1133+432 from the VLBI data at 2.3, 5.0, and 8.4~GHz}
      \label{1133+432_spec}
   \end{figure}

This is an empty field ($>23 ~mag$) in the optical (Stickel \&
K\"uhr, 1994). The 5~GHz VLBI image (Fig.~\ref{1133+432_i}) shows
a double source. The double structure exhibits two opposite
edge-brightened hotspot/lobes with spectral indices of 1.1 and 1.4
between 2.3 and 5~GHz for the northern and southern lobe
respectively, see Fig.~\ref{1133+432_spec}. Orienti et al (2004)
imaged the source with the VLBA at 5~GHz and they found a source
structure very similar to our Fig.~\ref{1133+432_i}. We confirm
that the source can be classified as a CSO. About 92\% of the
total flux density has been recovered in our 5~GHz image. No
polarization could be detected for this source.

\subsection{\object {1333+589} (\object {VCS1 J1335+5844}, \object {4C +58.26})}

\begin{figure}
      \includegraphics[width=7cm]{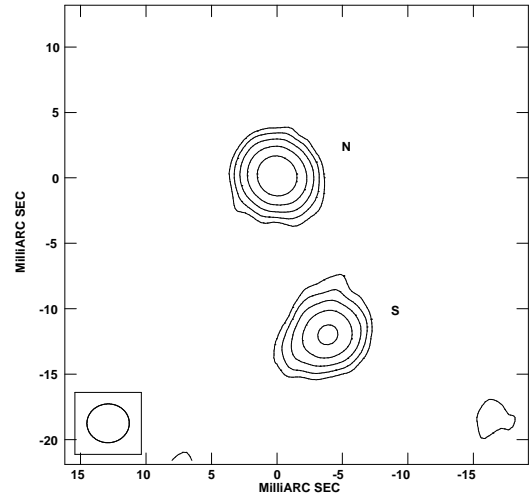}
      \caption{1333+589 at 5~GHz, the restoring beam is $3.2\times3.0$
        mas in PA $-87^{\circ}$,  the contours are 10 mJy/beam times -2, 1, 2, 4, 8, 16,
        with the peak of 290 mJy/beam.}
      \label{1333+589_i}
   \end{figure}

\begin{figure}
      \includegraphics[width=8cm]{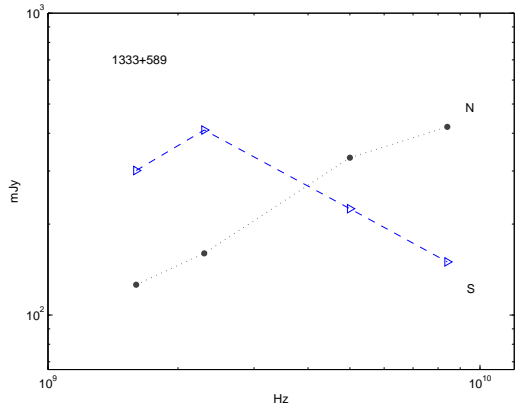}
      \caption{Component spectra of source 1333+589 from the VLBI data at 1.6, 2.3, 5.0, and 8.4~GHz}
      \label{1333+589_spec}
   \end{figure}

It is an empty field ($>23 ~mag$) in the optical (Stickel \&
K\"uhr, 1994). The radio spectrum, typical of a GPS source, peaks
at 4.9~GHz; the 5~GHz VLBI image shows a double structure
(Fig.~\ref{1333+589_i}). Images at 2.3 and 8.4 GHz are available
in paper I and the VLBA Calibrator Survey (VCS, part1, Beasley et
al. 2002). The northern component appears more compact than the
southern one, it may harbor a flat spectrum nucleus and it seems a
jet is embedded in it; its flux density is the same as Xu et al.
(1995) at 5~GHz. However, the southern component has a flux
density of 225 mJy in our observation, about 30\% less than in
Xu et al. (1995), the deficit may be due to poor uv-coverage because
the source has only 6 observing scans which are less than those
for other target sources in our observations. The distance between
the components `N' and `S' is 12.7 mas, the same as in Xu et al.
(1995). Both components have an overall radio spectrum with the
characteristic convex shape of self-absorbed regions
(Fig.~\ref{1333+589_spec}), and they can be interpreted in terms
of the lobes of the source. The spectral shape is confirmed in
more detail by higher frequency data (Orienti et al. 2006).

About 87\% of the Effelsberg flux density has been accounted for in
the VLBI image. No polarization is detected in the 5~GHz VLBI
observation.

\subsection{\object {1404+286} (\object {OQ~208}, \object {Mrk~0668}, \object {NVSS~J140700+282714})}

\begin{figure}
      \includegraphics[width=7cm]{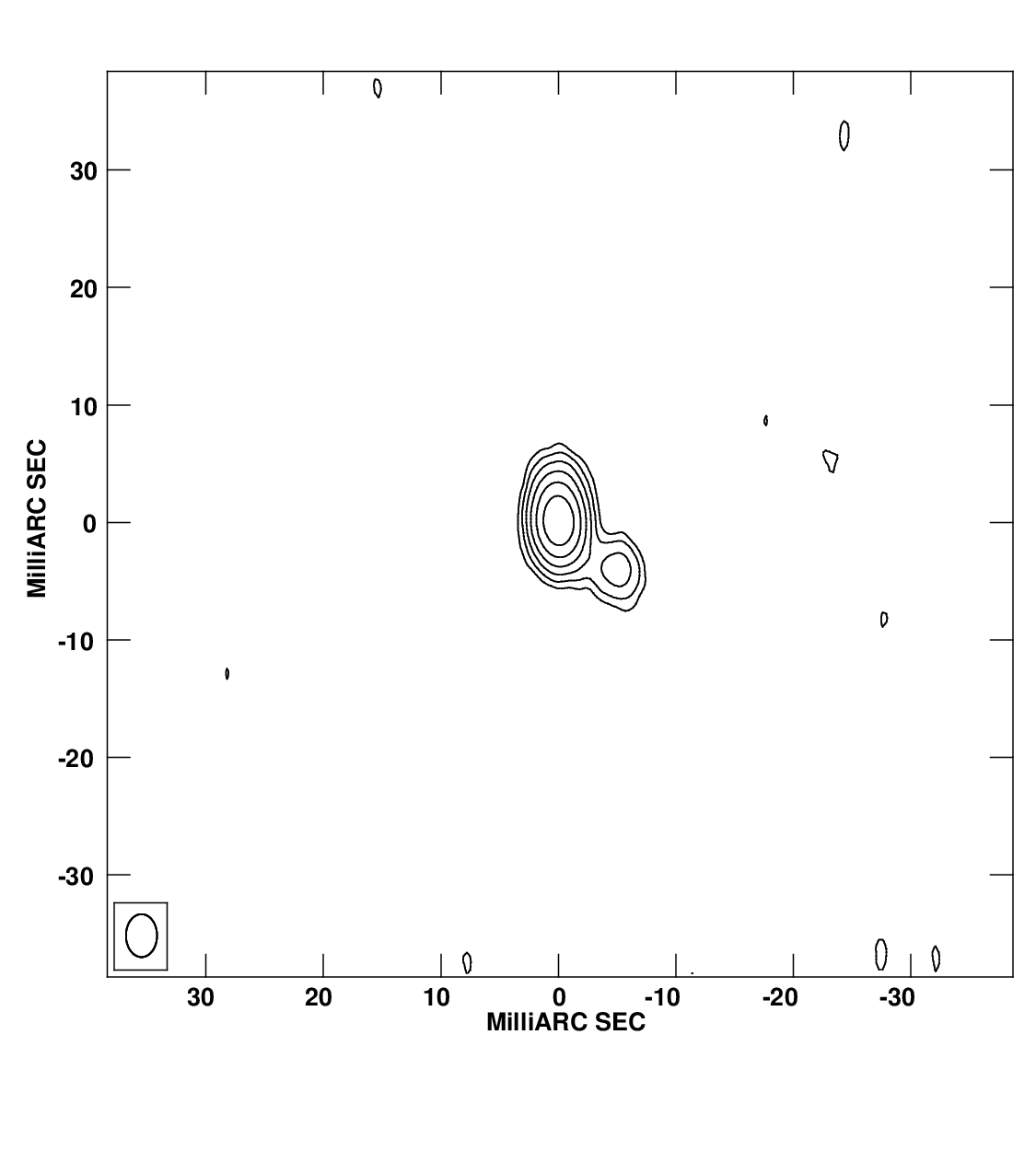}
      \caption{OQ208 at 5~GHz, the restoring beam is $3.7\times2.6$
        mas in PA $0^{\circ}$, the contours are 20 mJy/beam times -2, 1, 2, 4, 8, 16, 32, 64, with the peak
        of 1749 mJy/beam.}
      \label{oq208_i}
   \end{figure}

Our image  (Fig.~\ref{oq208_i}) shows a structure  which is
similar to previous VLBI observations at 5~GHz. This is a CSO as
confirmed by Stanghellini et al. (1998, 2000) and Lister et al.
(2003). No polarization is detected in our 5~GHz VLBI observation.
It was previously reported that its linear polarization is less
than $0.2\%$ at 5~GHz (Stanghellini et al.\ 1998). The source has
been found to be a Compton-thick AGN in X-rays (Guainazzi et
al.\ 2004); Yang \& Liu (2005) suggest the presence of a dense
medium capable of confining the radio source. The (ionized) medium
may have then depolarized the radio emission.

\subsection{\object {1433$-$040} (\object {PKS 1433$-$04}, \object {J1435$-$0414})}

\begin{figure}
      \includegraphics[width=7cm]{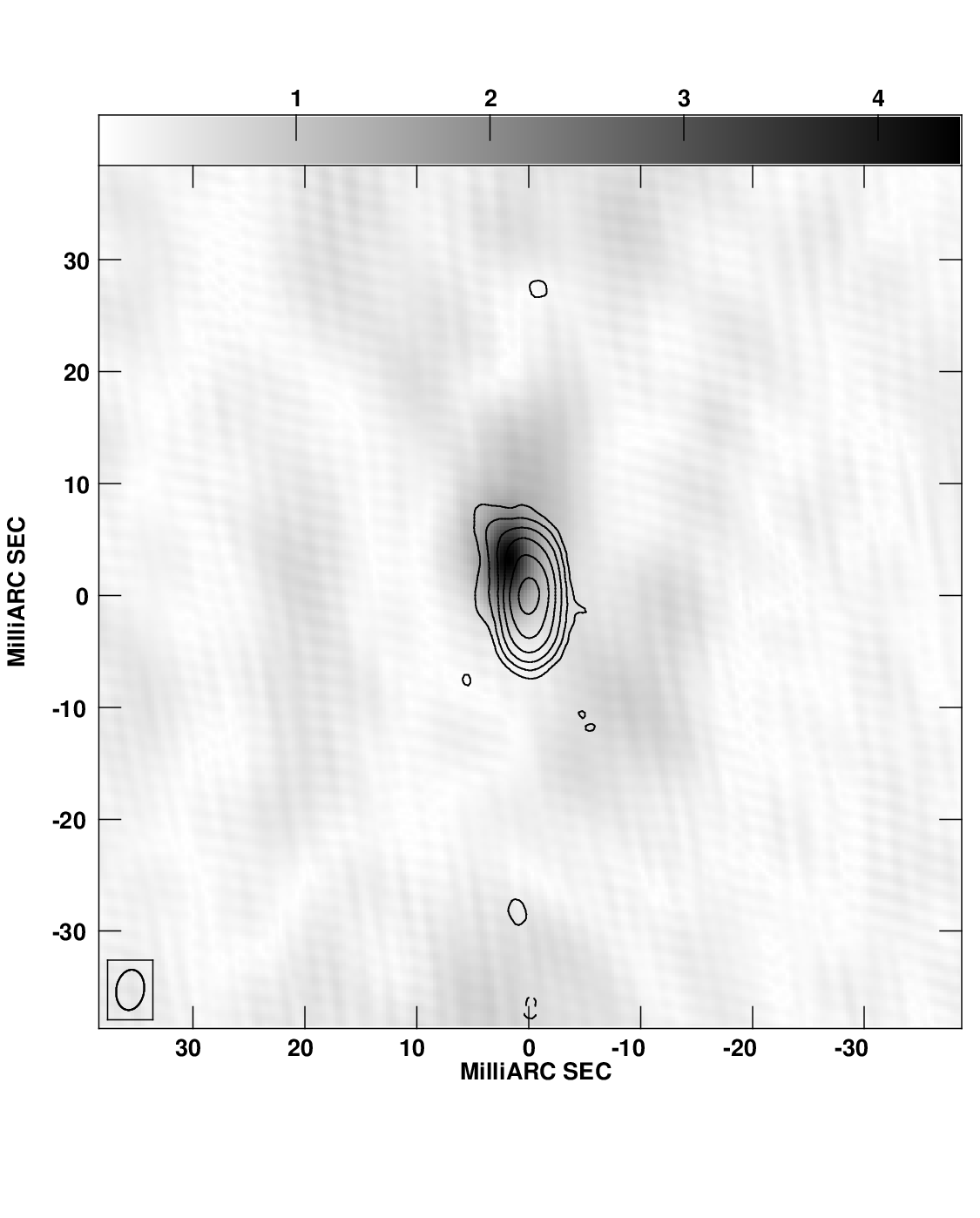}
\caption{1433$-$040 at 5~GHz, the restoring beam is $3.6\times2.5$
mas in PA $-8.6^{\circ}$, the contours are total intensity with
2.0 mJy/beam times -2, 1, 2, 4, 8, 16, 32, 64, 90, and the peak of
84.2 mJy/beam. The overlay in grey is linearly polarized
intensity, with the peak of 3.8 mJy/beam.}
      \label{1433-040_ip}
   \end{figure}

The source is identified as a galaxy with $m_{r}$=22.3
(Stanghellini et al.\ 1993). It can be found in O'Dea (1996)
and in the early GPS sample list by Spoelstra (1985). The parsec
scale morphology appears as a core-jet (Fig.~\ref{1433-040_ip}),
and the 2.3~GHz VLBI image from the VCS3 (VLBA calibrator survey
3, Petrov et al. 2005) clearly shows a core-jet structure oriented
to the North-East. The 5 GHz image can be fitted with two Gaussian
components, a central compact one, likely to be the core, and a
weaker and broader one to the North-East with a separation of 1.4
mas and $PA$=12.4$^{\circ}$. The core-jet classification is
supported by the positive detection of polarization in this
source. Integrated polarization of 3.6\% is detected in the 5~GHz
VLBI observation, with an EVPA of 143$^{\circ}$
(Fig.~\ref{1433-040_p}). The polarization seems largely to come
from the jet component because the polarized emission lies close
to the jet position. We obtained a fractional polarization of
3.8\% with an EVPA of 134$^{\circ}$ in the WSRT data for this
source.

The radio spectrum of the source turns over around 1~GHz. The
total flux density has increased from 200 mJy (observed 1980,
Parkes Catalogue 1990) to 246 mJy in this observation. About 83\%
of the total flux density has been detected in the 5~GHz VLBI
image.

\subsection{\object {1509+054} (\object {VCS2 1511+0518}, \object {JVAS J1511+0518})}

\begin{figure}
      \includegraphics[width=7cm]{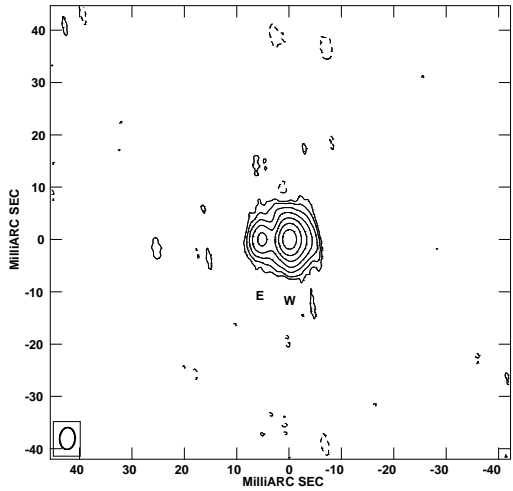}
      \caption{1509+054 at 5~GHz using uniform weighting with an AIPS
      robust parameter of 1, the restoring beam is $4.2\times2.94$
        mas in PA $-3.3^{\circ}$, the contours are 4 mJy/beam times -2, -1, 1,
        2, 4, 8, 16, 32, 64, with the
        peak of 407 mJy/beam.}
      \label{1509+054_i}
   \end{figure}

This object has been identified with a Seyfert-1 galaxy at a
redshift of 0.084 (Chavushyan et al. 2001). The 5~GHz VLBI image
shows an asymmetric double (Fig.~\ref{1509+054_i}), which is
similar to the 8.4~GHz image in paper I. The eastern component is
more compact than the western one which is marginally resolved in
(Fig.~\ref{1509+054_i}). The spectral indices of the eastern
component and western part are -1.96 and 0.61 between 5 and
8.4~GHz (from Fig.~\ref{1509+054_i} at 5~GHz and paper I at
8.4~GHz). From the compactness and steep rising spectrum, the
component `A' may harbor the core of the source. However, if we
consider also the VCS data and higher frequency VLBA observations
by Orienti et al. (2006), we can also interpret the source as
dominated by two self-absorbed, micro hot-spots/lobes with
slightly different turnover frequency.

\begin{figure}
      \includegraphics[width=7cm]{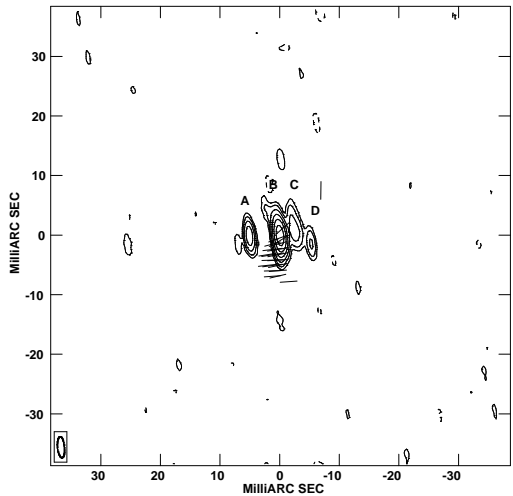}
      \caption{1509+054 at 5~GHz using uniform weighting with an AIPS robust
      parameter of $-$2, the restoring beam is $3.6\times1.2$ mas in PA
      $5.2^{\circ}$, the contours are 6 mJy/beam times -2, -1, 1, 2, 4, 8, 16,
      32, with the peak of 325 mJy/beam. The electric vectors of
      possible polarizations are superimposed.}
      \label{1509+054_i2}
   \end{figure}

The total flux density of the source has increased from 526 mJy
(Dallacasa et al. 2000) to 688 mJy in this observation, suggesting
it is variable and this is a core-jet source. Weak polarization is
possibly detected in the central component
(Fig.~\ref{1509+054_i2}) in our 5~GHz VLBI image; we use it as an
upper limit in Table~\ref{table:ex004}, although the EVPA differs
from that of the WSRT result. The source may be associated with
the X-ray source \object {1WGA J1511.6+0518}.

\subsection{\object {1518+046} (\object {VCS1 J1521+0430}, \object {4C +04.51})}

\begin{figure}
      \includegraphics[width=7cm]{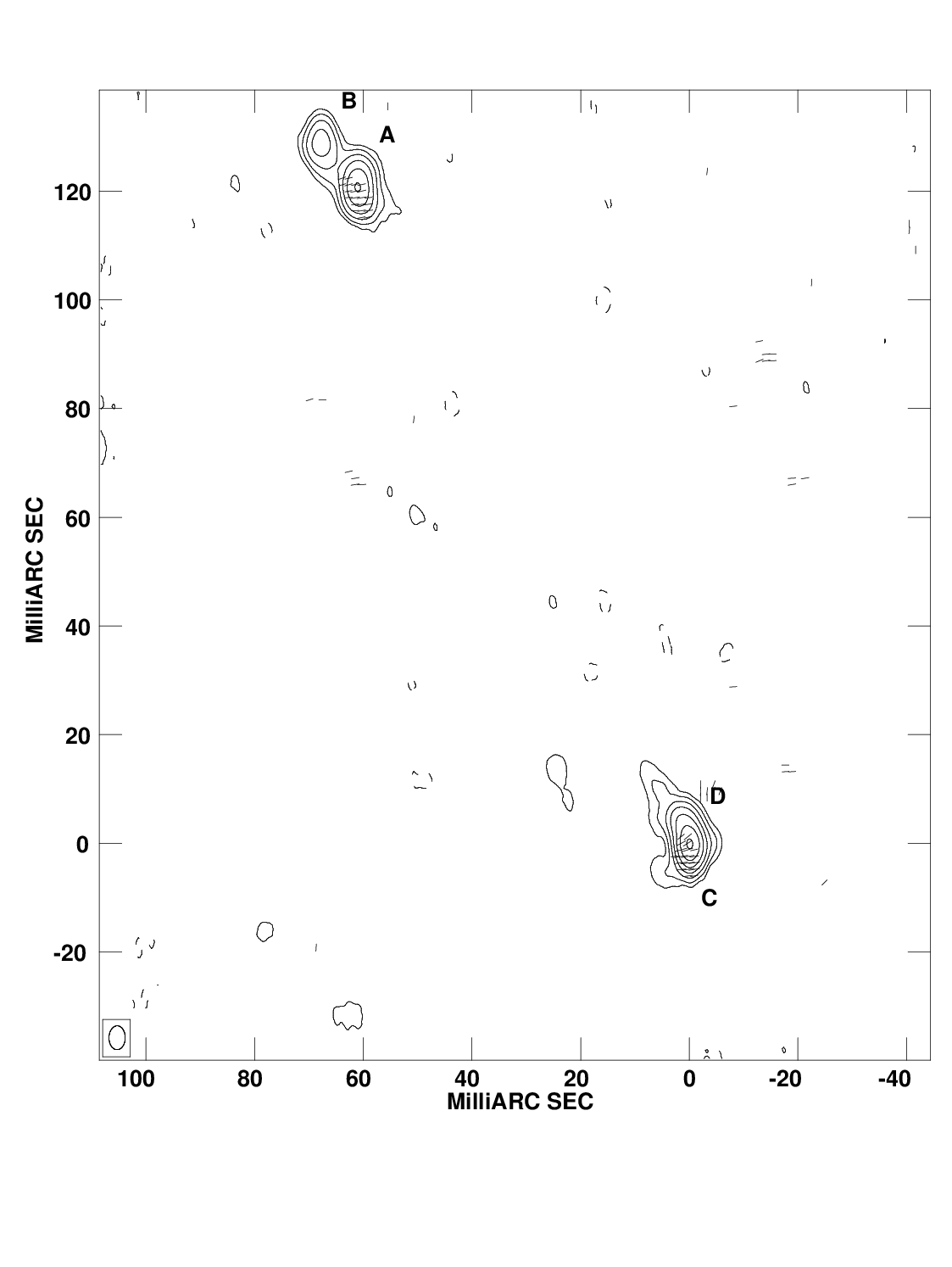}
      \caption{1518+046 at 5~GHz, the restoring beam is $4.4\times3.0$ mas in
      PA $-0.3^{\circ}$, the contours are 4 mJy/beam times
      -2, -1, 1, 2, 4, 8, 16, 32, 64, with the peak of 276 mJy/beam. The electric
      vectors of possible polarizations are superimposed.}
      \label{1518+046_i}
   \end{figure}

The 5~GHz VLBI image shows a classical double
(Fig.~\ref{1518+046_i}), with two hotspots in the northern and
tail/lobe in the southern part. The identification of these
components as hotspots/lobes is supported by the spectra, as given
in paper I (see also Dallacasa et al. 1998). The total flux
density is stable from 1.03 Jy (Parkes Catalogue 1990) to 1.06 Jy
in this observation. We conclude that this is a CSO or, possibly a
MSO, given its size around 1~kpc. Possible polarization is
detected at the hotspots in the 5~GHz image, we use it as an upper
limit in Table~\ref{table:ex004}, although the EVPA differs from
that of the WSRT result.

\subsection{\object {1751+278} (\object {MG2 J175301+2750}, \object {JVAS 1753+2750})}

\begin{figure}
      \includegraphics[width=7cm]{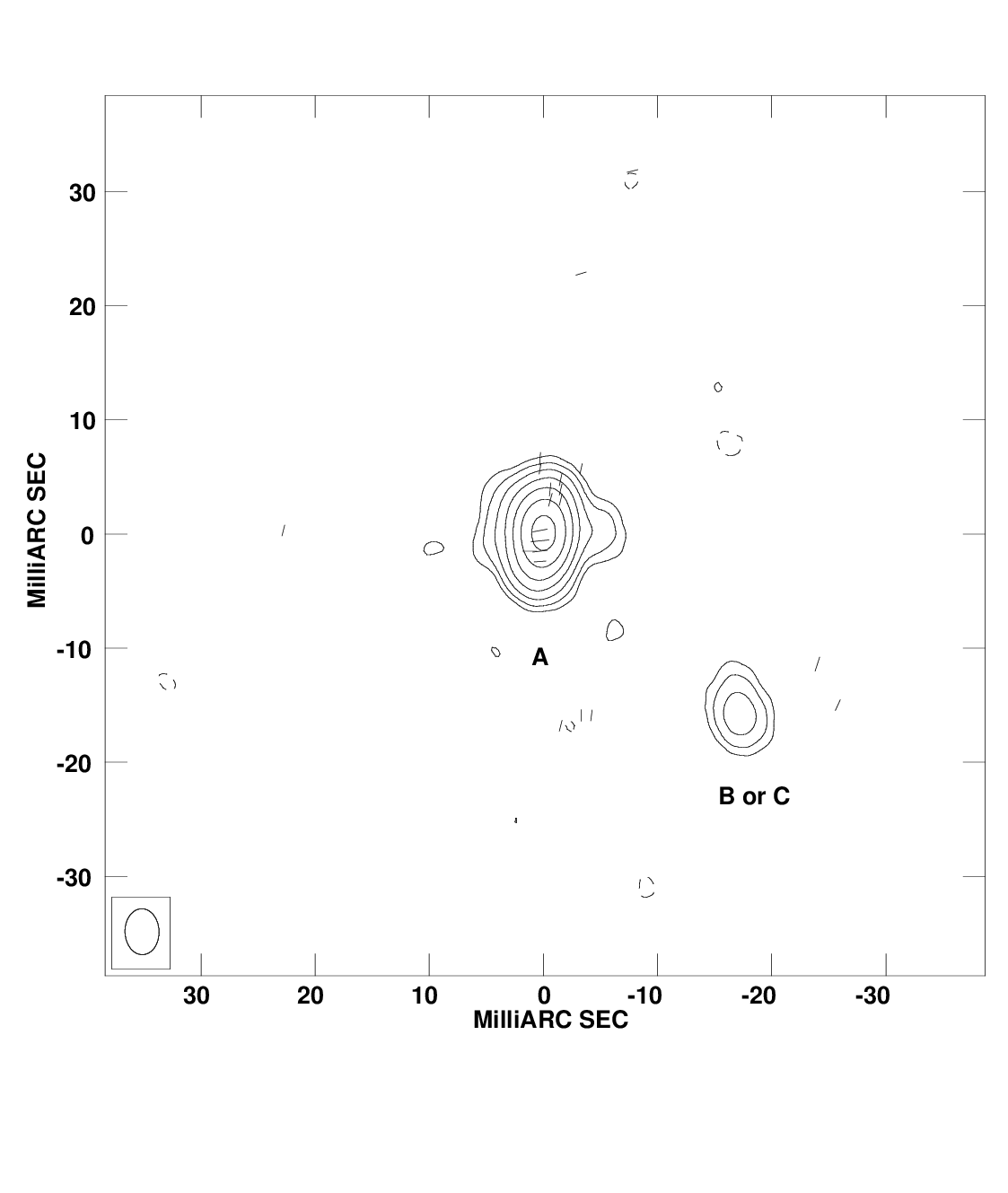}
      \caption{1751+278 at 5~GHz, the restoring beam is $4.1\times3.0$
        mas in PA $1.8^{\circ}$, the contours are 2.0 mJy/beam times -2, -1, 1, 2, 4, 8, 16, 32, 64,
        with the peak of 170 mJy/beam.
        The electric vectors of possible polarizations are superimposed.}
      \label{1751+278_i}
   \end{figure}

The 5~GHz VLBI image exhibits asymmetric double structure. The
northern component `A' is about 10 times brighter than the
southern one (Fig.~\ref{1751+278_i}). We derive a spectral index
of 0.59 for component `A' between 1.6 and 5~GHz. We could relate
the weaker one at 5 GHz to component `B' or `C' or even both at
1.6 GHz because its separation of 23.3 mas from component `A' at 5
GHz, is close both to the 22 mas separation of `B' from `A' and to
the 23 mas separation of `C' from `A' (paper I) at 1.6 GHz. If it
is `B' its spectral index is 0.6, if it is `C' then $\alpha=-0.4$,
or it is a combination of `B' and `C' in which case $\alpha=0.8$.
Therefore, if the weaker component at 5 GHz is `B' or `B+C' at 1.6
GHz, the steep spectrum would suggest the source has a double
lobe/hotspots; if it is `C' then the resultant rising spectrum
would indicate `C' is the core of the source. The source is
suggested to vary at 1.4 GHz, and may not be a genuine GPS source
(paper I). At 5 GHz its total flux density has slightly changed
from 280 mJy (Griffith et al. 1990) to 260 mJy in this
observation, although this difference is consistent with no
variability within 1 $\sigma$. However, further VLBI observations
at other frequencies would be required to clarify the source
classification. The total flux density at 5~GHz is recovered in
the VLBI image. Possible polarization is detected at the strong
component in the 5~GHz image, which we use as an upper limit in
Table~\ref{table:ex004}.

\subsection{\object {1824+271} (\object {MG2 J182632+2707})}

\begin{figure}
      \includegraphics[width=7cm]{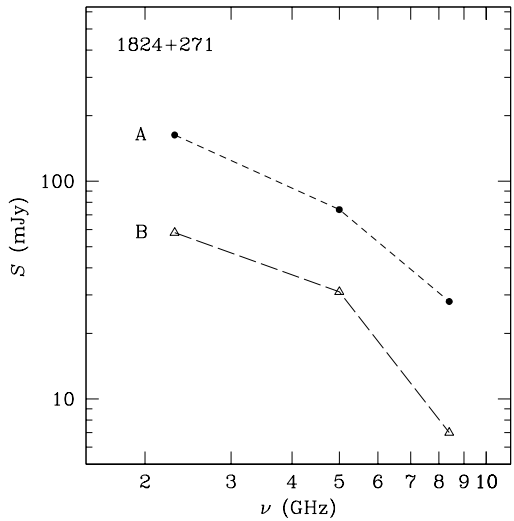}
      \caption{Component spectra of source 1824+271 from the VLBI data at 2.3, 5.0, and 8.4~GHz}
      \label{1824+271_spec}
   \end{figure}

The 5~GHz VLBI image (Fig.~\ref{1824+271_i}) shows a symmetric
double structure, which is similar to 2.3/8.4~GHz images (paper
II). The spectral indices for components `A' and `B' respectively
are 1.0 and 0.8 between 2.3~GHz and 5~GHz while between 5~GHz and
8.4~GHz they both become steeper than $\alpha=1.0$
(Fig.~\ref{1824+271_spec}), suggesting that the source structure
can be interpreted in terms of a double-lobed object. We then
confirm that the source is a CSO. The total flux density has
slightly increased from 98 mJy (Griffith et al. 1990) to 111 mJy
in this observation, although there is consistency with no flux
density variability within 1 $\sigma$. The whole flux density has
been accounted for in the VLBI image. No polarization is detected
in the 5~GHz VLBI and WSRT observations.

\subsection{\object {2121$-$014} (\object {PKS 2121$-$01}, \object {NVSS J212339$-$011234})}

The 5~GHz VLBI image shows two bright regions, likely the
lobes, and an additional weaker feature, possibly the  core or a
knot in a jet, between them (Fig.~\ref{2121-014_i}). We estimated
the spectral index of the core candidate between 2.3~GHz and 5~GHz
by using the peak flux density of 45 mJy at 2.3~GHz and the measure
from our data at 5~GHz, and it turns out to be very steep $\alpha=1.5$
(details are explained further down), unless dramatic flux density
variability took place. In that case, it is more likely a fading knot in a
jet or part of the tailed structure characteristic of the radio
lobes, which is also consistent with the non detection at 8.4~GHz
(paper II). In paper II, we fitted two Gaussian components `A' and
`B' in one `tvwindow' using the AIPS task JMFIT resulting in
components `A' and `B' largely overlapping with a separation of 4
mas only (the value of separation is not listed in paper II). Here
we refit components `A' and `B' at 2.3~GHz with single components
in separate windows, which gives peak/beam and integrated flux
density 275, 330 mJy for `A' and 45, 120 mJy for `B' respectively,
with a separation of 16.0 mas between them which is consistent
with the separation 16.8 mas measured at 5~GHz. With the
remeasured data at 2.3~GHz, the spectral indices between 2.3~GHz
and 5~GHz are 1.0, 0.7 for `C', `A' (we have combined components
`C1'/`C2' into `C', `A1'/`A2' into `A' at 5~GHz).

\begin{figure}
      \includegraphics[width=7cm]{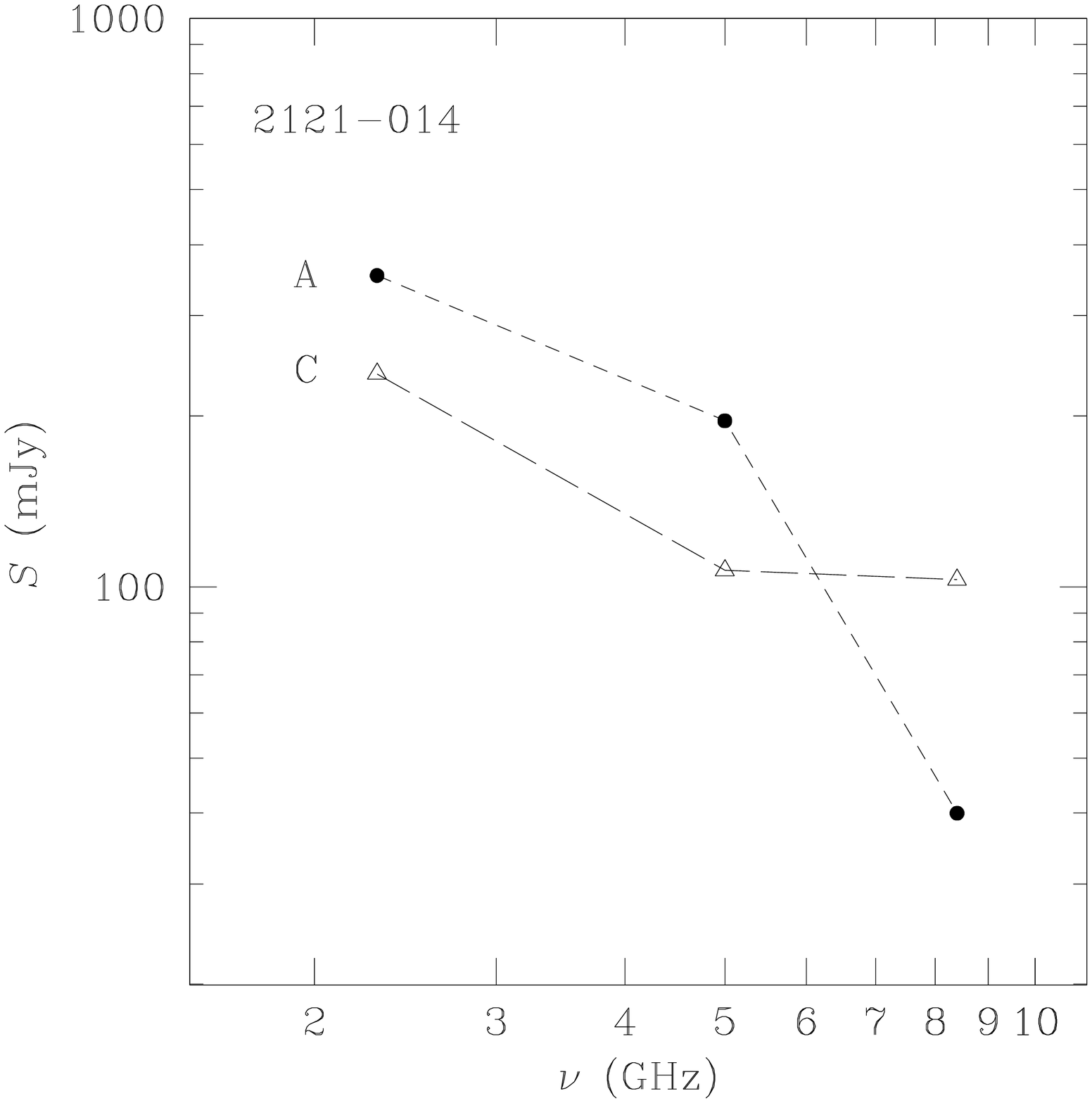}
      \caption{Component spectra of source 2121$-$014 from the VLBI data at 2.3, 5.0, and 8.4~GHz}
      \label{2121-014_spec}
   \end{figure}

For $\alpha=1.5$ of the `B' component between 2.3 and 5 GHz, we
have used the peak/beam flux instead of integrated flux at 2.3
GHz, with a comparable angular size to component `B' at 5 GHz.
From the double lobe-like features and their steep spectra
(Fig.~\ref{2121-014_spec}) we confirm the source is a CSO,
although the `central' component `B' has not been confirmed as a
core. The total flux density seems stable from 320 mJy (Parkes
Catalogue 1990) to 327 mJy in this observation. No polarization is
detected in the 5~GHz VLBI observation, and the WSRT value is
similarly low.

\subsection{\object {PKS 2322$-$040} (\object {NVSS J232510$-$034446})}

\begin{figure}
      \includegraphics[width=7cm]{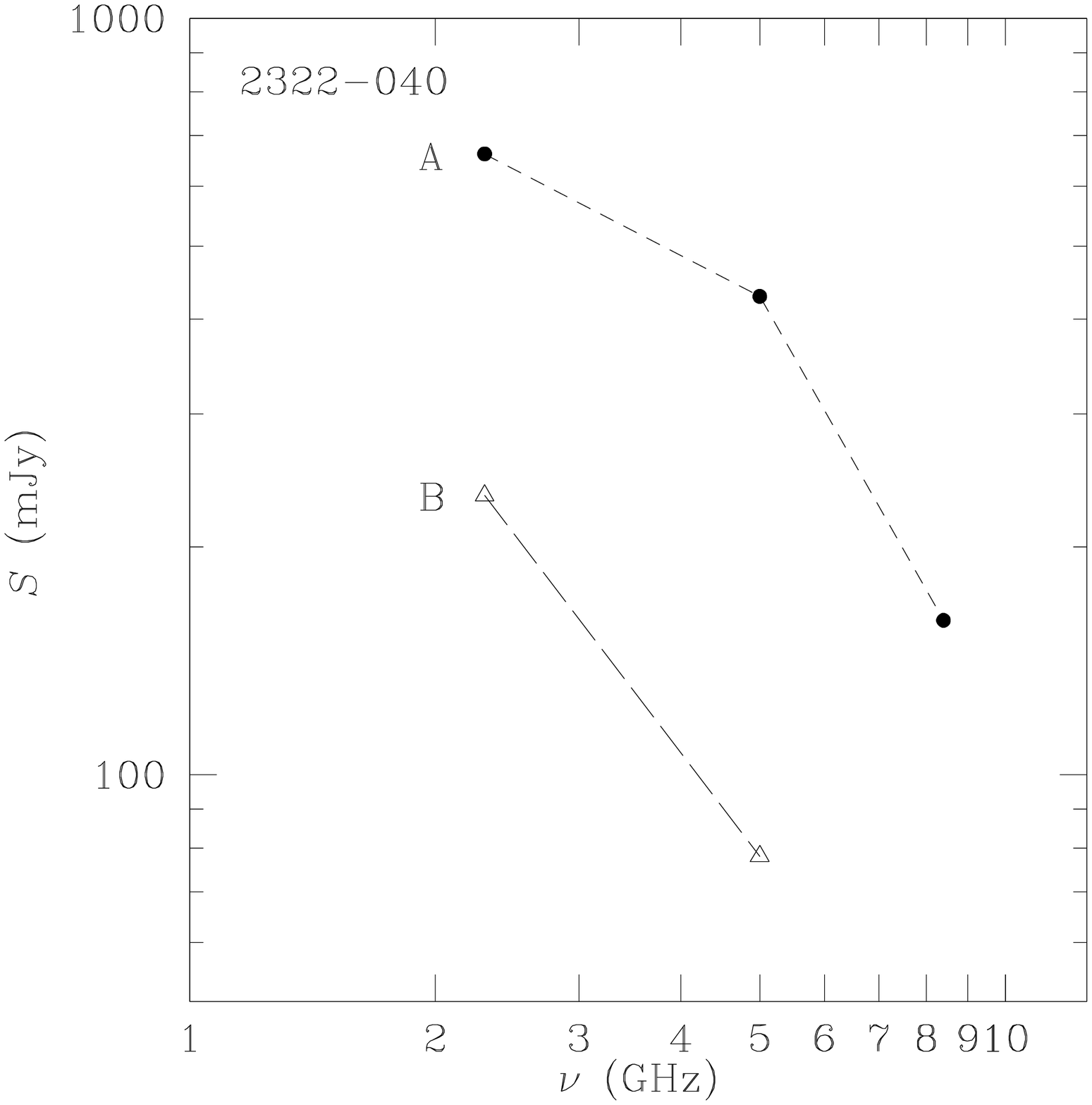}
      \caption{Component spectra of source 2322$-$040 from the VLBI data at 2.3, 5.0, and 8.4~GHz}
      \label{2322-040_spec}
   \end{figure}

With a typical GPS spectrum peaked at 1.4 GHz, the 5~GHz VLBI
image exhibits double lobe/hotspots and a weak jet/tail-like
emission to the northern component (Fig.~\ref{2322-040_i}). It
seems similar to that at 2.3~GHz (paper II) where a sign of the
jet might be embedded in component `A' but not well
resolved. Component `B' assumes a lobe-like morphology in our
image at 5~GHz. The spectral indices of components `A', `B' are
0.56, 1.41 respectively between 2.3 and 5~GHz (see also
Fig.~\ref{2322-040_spec}). The total flux density has increased by
10\% from 500 mJy (Parkes Catalogue 1990) to 548 mJy in this
observation. Based on the steep spectra and the radio structure,
we tentatively classify the source as a CSO. No polarization is
detected either in the 5~GHz VLBI or the WSRT observations.

\subsection{\object {2323+790} (\object {NVSS J232503+791715})}

This object is classified as a galaxy (Stickel \& K\"uhr 1996). On
the large scale, the NVSS image exhibits two components separated
by $\sim 2$ arcmin.  It is not clear whether they are separate
sources. The WENSS map indicates the source has multiple
components. On the very small scale, the 5~GHz VLBI image shows
two components, both resolved; the easternmost is the brightest
and has a North-South elongation, which is quite perpendicular to
the major axis of the weaker western component. The overall
structure can be classified as a double (Fig.~\ref{2323+790_i}).
This is the first VLBI image for the source and we cannot provide
a proper classification on its basis only. It belongs to the S5
catalogue, and the total flux density seems stable at 5~GHz from
448 mJy (S5 data, observed in 1978, K\"uhr et al. 1981) to 438 mJy
in this observation. We consider this source as a CSO candidate,
requiring further observations for a proper classification. About
93\% of the total flux density has been detected in the VLBI
image. No polarization is detected either in the 5~GHz VLBI or in
the WSRT observation.

\section{Discussion}

CSOs are compact radio sources with relatively steep spectrum
double lobes on the opposite sides of the core. Ideally, a core
component with a spectrum flatter than the lobes must be
identified before sources can be claimed to be a CSO. For some
CSOs with a jet axis very close to the plane of the sky the core
may be so weak as to be undetectable unless very high dynamic
range images are available; yet a CSO identification can still be
secure if there are quite symmetric, edge-brightened lobes (Taylor
\& Peck 2003). Examples are the radio sources \object{0914+114},
\object{1133+432}, \object{1518+046}, \object{1824+271},
\object{2121$-$014} presented in this paper, for which the CSO
classification can be proposed despite the failure to detect the
core. These sources have resolved, steep spectrum components
dominating the pc-scale radio emission. The long term ($>15$
years) variations as indicated in the last section are $-16\%$,
unknown, $+3\%$, $+12\%$, and $+2\%$ respectively for the 5 target
sources classified as CSOs. During the 2.4 years that elapsed
between the present and our earlier observations their variations
should not affect substantially their spectral indices and then
their CSO classification can be considered secure.

The optical counterparts of the 14 sources in our sample are
primarily galaxies (10), with the addition of 2 quasars and 2
empty fields. One quasar (\object{DA193}) and one galaxy
(\object{1433$-$040}) show significant linear polarizations
$\geq1\%$, while there is marginal detection ($\leq0.5\%$) of
polarized emission in the remaining quasar (\object{1518+046}) and
2 further galaxies (\object{1509+054} and \object{1751+278}). The
remaining sources do not show any polarization.

These results are consistent with earlier work on the GPS sources,
known to show very low or no polarization. They may have been
largely depolarized by a dense ionized ambient medium
characterized by small scale inhomogeneities in its magnetic field
and in which the radio source is embedded. An example of such a medium
can be found in \object{DA193}, known to have a very high
rest-frame rotation measure ($>4700\;\rm rad\,m^{-2}$; Lister \&
Marscher 1998), possibly responsible for some amount of Faraday
depolarization as well. A similarly dense ionized medium
surrounding the Compton-thick \object{OQ208} (Guainazzi et al.
2004) may have led to undetectable radio polarization.

Given that in general the rotation measure is not or poorly known
for GPS sources, no correction for the effect of Faraday rotation
on intrinsic orientation of the electric vector ($\chi$) has been
applied in the images presented here, and therefore the intrinsic
magnetic field vectors are not necessarily perpendicular to the
{\it observed} electric vectors shown in the images.

\section{Summary and Conclusions}

\begin{enumerate}

\item We have obtained 5~GHz total intensity VLBI images for 14
GPS sources. The parameters of source structure and spectra have
been derived, allowing, for a number of cases, a secure
morphological classification.

\item Two core-jet sources (\object{1433$-$040} and
\object{DA193})
  show integrated fractional polarization of 3.6\% and 1.0\%
  respectively. Three sources show possible weak polarization $\leq
  0.5\%$, while the remaining nine sources are found to be unpolarized
  at 5~GHz, confirming that this is a characteristic property of GPS
  sources.

\item The sources \object{1133+432}, \object{1824+271} and
2121$-$014
  have been confirmed as CSOs, and three new CSOs have been classified
  as such on the basis of new evidence coming from the 5~GHz
  images and the spectral indices: they are 0914+114,
  \object{1518+046}, and \object{2322$-$040}.

\item The sources \object{1333+589}, \object{1751+278} and
  \object{2323+790} which show double structure are likely candidate CSOs or
  core-jet sources; further VLBI observations are required for their
  proper classification. \object{0554$-$026}, \object{1433$-$040} and
  \object{1509+054} are classified as core-jet sources.

\item The jet component in the quasar \object{DA193} has a
  superluminal motion of $3.3\pm0.6\, h^{-1}c$ in 5.5 years,
  confirming earlier results.

\end{enumerate}

\begin{acknowledgements}

LX is grateful for support from the KNAW/CAS (Royal Dutch
Academy of Sciences/Chinese Academy of Sciences) bilateral
agreement. We thank Dave Graham for the total flux density
measurement at Effelsberg. The European VLBI Network is a joint
facility of European, Chinese, South African and other radio
astronomy institutes funded by their national research councils.
The WSRT is operated by ASTRON with financial support from the
Netherlands Organization for Scientific Research (NWO). This
research has made use of the NASA/IPAC Extragalactic Database
(NED) which is operated by the Jet Propulsion Laboratory, Caltech,
under contract with NASA. Parkes Catalogue (1990) was made by
Australia Telescope National Facility.

\end{acknowledgements}

\clearpage

\begin{figure}
      \includegraphics[width=7cm]{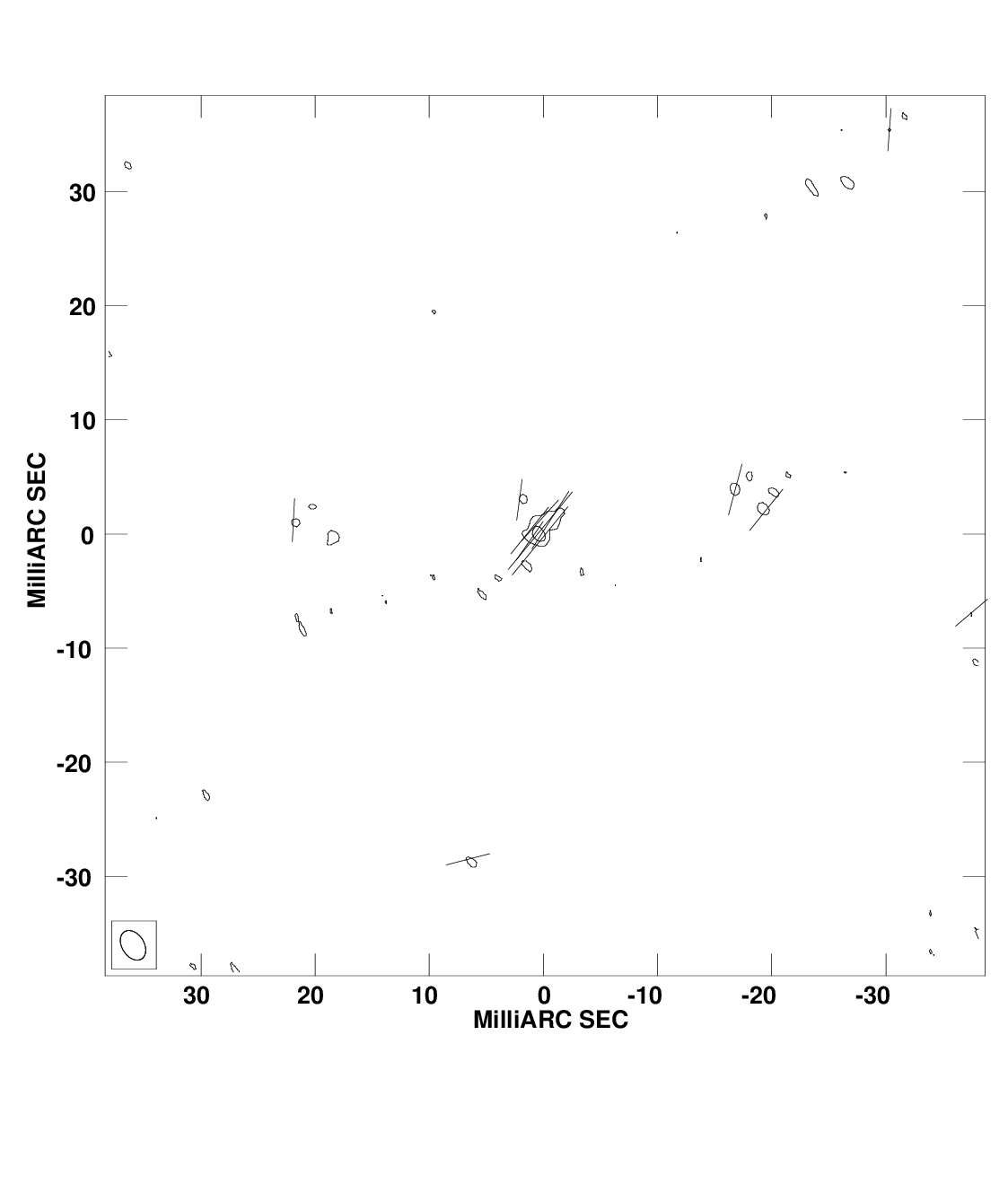}
      \caption{Linear polarization map of DA193 with polarization angle superimposed
      at 5~GHz, the restoring beam is $2.8\times1.9$ mas
        in PA $32.5^{\circ}$, the contours are 20 mJy/beam times -2, -1, 1, 2, with the peak
        of 56 mJy/beam.}
      \label{da193_p}
   \end{figure}

\begin{figure}
      \includegraphics[width=7cm]{3885fig3}
      \caption{0554$-$026 at 5~GHz, the restoring beam is
        $3.8\times2.5$ mas in PA $-5.8^{\circ}$,  the contours are
        1.5 mJy/beam times -2, -1, 1, 2, 4, 8, 16, 32, with the peak of 71 mJy/beam.
        The electric vectors of possible polarizations are superimposed.}
      \label{0554-026_i}
   \end{figure}

\clearpage

\begin{figure}
      \includegraphics[width=7cm]{3885fig6}
 \caption{1133+432 at 5~GHz, the restoring beam is $2.2\times2.0$ mas
   in PA $51^{\circ}$,  the contours are 1.5 mJy/beam times -2, -1, 1, 2, 4, 8, 16, 32, 64, with
   the peak of 134 mJy/beam.
   }
      \label{1133+432_i}
   \end{figure}

\clearpage

\begin{figure}
      \includegraphics[width=7cm]{3885fg12}
      \caption{Linear polarization map of 1433$-$040 with polarization angle superimposed at 5~GHz,
      the restoring beam is $5.0\times3.4$
        mas in PA $1.8^{\circ}$,  the contours are 0.5 mJy/beam times -2, -1,
        1, 2, 4, with the peak of 3.8 mJy/beam.}
      \label{1433-040_p}
   \end{figure}

\clearpage

\clearpage

\begin{figure}
      \includegraphics[width=7cm]{3885fg17}
      \caption{1824+271 at 5~GHz, the restoring beam is $4.6\times3.5$
        mas in PA $1.7^{\circ}$, the contours are 0.9 mJy/beam times -2, 1, 2, 4, 8, 16, 32, with the peak
        of 50 mJy/beam.}
      \label{1824+271_i}
   \end{figure}

\begin{figure}
      \includegraphics[width=7cm]{3885fg19}
      \caption{2121$-$014 at 5~GHz, the restoring beam is $4.7\times3.4$
        mas in PA $4.9^{\circ}$,  the contours are 2 mJy/beam times -2, 1, 2, 4, 8, 16, 32, with the peak
        of 107 mJy/beam.}
      \label{2121-014_i}
   \end{figure}

   \clearpage

\begin{figure}
      \includegraphics[width=7cm]{3885fg21}
      \caption{2322$-$040 at 5~GHz, the restoring beam is $4.0\times2.3$
        mas in PA $-7.6^{\circ}$,  the contours are 6 mJy/beam times -2, 1, 2,
        4, 8, 16, 32, with the peak
        of 299 mJy/beam.}
      \label{2322-040_i}
   \end{figure}

\begin{figure}
      \includegraphics[width=7cm]{3885fg23}
      \caption{2323+790 at 5~GHz, the restoring beam is $3.8\times3.1$
        mas in PA $-1.6^{\circ}$, the contours are 2 mJy/beam times -2, 1, 2, 4, 8, 16, 32, 64, with the peak
        of 184 mJy/beam.}
      \label{2323+790_i}
   \end{figure}



\begin{thebibliography}{}



\bibitem[]{}
Beasley, A.J., D. Gordron, A.B. Peck, et al. 2002, ApJS 141, 13


\bibitem[]{}

Chavushyan, V., Mujica, R., Gorshkov, A. G., et al., 2001,
Astronomy Report, Vol.45, No.2, 79


\bibitem[]{}
Cotton, W. D., Dallacasa, D., Fanti, C., Fanti, R., Foley, A. R.,
Schilizzi, R. T., Spencer, R. E., 2003, A\&A 406, 43


\bibitem[]{}

de Vries, W. H., Barthel, P. D., O'Dea, C. P., 1997, A\&A 321, 105


\bibitem[]{}
Dallacasa, D., Bondi, M., Alef, W., Mantovani, F., 1998, A\&AS
129, 219


\bibitem[]{}

Dallacasa, D., Stanghellini, C., Centonza, M., Fanti, R., 2000,
A\&A 363, 887


\bibitem[]{}

Dallacasa, D., 2004, in the Proceedings of 7th European VLBI
Network Symposium, eds: Bachiller, R., Colomer, F., Desmurs, J.
F., de Vicente, P., p59


\bibitem[]{}

Fanti, C., Fanti, R., Dallacasa, D., Schilizzi, R. T., Spencer, R.

E., Stanghellini, C., 1995, A\&A 302, 317



\bibitem[]{}

Fanti, C., 2000, in the Proceedings of 5th European VLBI Network
Symposium, held at Chalmas Technical University, eds. A. G.
Polatidis, R. S. Booth, \& Y. Pihlstr\"om (Published by Onsala
Space Observatory), p73



\bibitem[]{}

Griffith, M.,  Langston, G., Heflin, M., Conner, S., Lehar, J.,
Burke, B., 1990, ApJS 74, 129



\bibitem[]{}

Guainazzi, M., Siemiginowska, A., Rodriguez-Pascual, P., and
Stanghellini, C., 2004, A\&A 421, 461



\bibitem[]{}

Gugliucci, N. E., Taylor, G. B., Peck, A. B., Giroletti, M., 2005,
ApJ 622, 136



\bibitem[]{}

Heckman, T. M., O'Dea, C. P., Baum, S. A., Laurikainen, E., 1994,
ApJ 428, 65



\bibitem[]{}

Homan, D. C., Attridge, J. M., Wardle, J. F. C, 2001, ApJ 556, 113



\bibitem[]{}

K\"uhr, H., Pauliny-Toth, I. I. K., Witzel, A., Schmidt, J., 1981,
AJ 86, 6



\bibitem[]{}

Lister, M. L., Marscher, A. P., 1998, ApJ 504, 702



\bibitem[]{}

Lister M. L., Kellermann K. I., Vermeulen R. C., Cohen M. H.,
Zensus J. A., Ros E., 2003, ApJ 584, 135L


\bibitem[]{}
Murgia, M., Fanti, C., Fanti, R., Gregorini, L., Klein, U., Mack,
K.-H., Vigotti, M., 1999, A\&A 345, 769

\bibitem[]{}

Murgia, M., 2003, PASA 20, 19



\bibitem[]{}

O'Dea, C. P., Stanghellini, C., Baum, S. A., Charlot, S., 1996,
ApJ 470, 806



\bibitem[]{}

O'Dea, C. P., 1998, PASP 110, 493

\bibitem[]{}
Orienti, M., Dallacasa, D., Tinti, S., Stanghellini, C., 2006,
A\&A in press


\bibitem[]{}

Orienti, M., Dallacasa, D., Fanti, C., Fanti, R., Tinti, S.,
Stanghellini, C., 2004, A\&A 426, 463


\bibitem[]{}
Owsianik, I., Conway, J. E., 1998, A\&A 337, 69


\bibitem[]{}
Pearson, T. J., Readhead, A. C. S., 1988, ApJ 328, 114



\bibitem[]{}

Petrov, L., Kovalev, Y. Y., Fomalont, E., Gordon D., 2005, AJ 129,
1163


\bibitem[]{}
Polatidis, A. G., Conway, J. E., 2003, PASA 20, 69


\bibitem[]{}

Scott, W. K., Fomalont, E. B., Horiuchi, S., Lovell, J. E. J., et
al., 2005, ApJS 155, 33



\bibitem[]{}
Snellen, I. A. G., Schilizzi, R. T., Miley, G. K., de Bruyn, A.
G., Bremer, M. N., R\"ottgering, H. J. A., 2000, MNRAS 319, 445



\bibitem[]{}
Stanghellini, C., O'Dea, C. P., Dallacasa, D., Cassaro, P., Baum,
S. A., Fanti, R., Fanti, C., 2005, A\&A 443, 891


\bibitem[]{}

Stanghellini, C., O'Dea, C. P., Baum, S. A., Laurikainen, E.,
1993, ApJS 88, 1



\bibitem[]{}

Stanghellini, C., O'Dea, C. P., Dallacasa, D., Baum, S. A., Fanti,
R., Fanti, C., 1998, A\&AS 131, 303



\bibitem[]{}

Stanghellini, C., Dallacasa, D., Bondi, M., Xiang, L., 2000, in
the Proceedings of 5th European VLBI Network Symposium, held at
Chalmas Technical University, eds. A. G. Polatidis, R. S. Booth,
\& Y. Pihlstr\"om (Published by Onsala Space Observatory), p99



\bibitem[]{}

Stickel, M., K\"uhr, H., 1994, A\&AS 103, 349



\bibitem[]{}

Stickel, M., K\"uhr, H., 1996, A\&AS 115, 1



\bibitem[]{}

Spoelstra, T. A. T., Patnaik, A. R., Gopal-Krishna, 1985, A\&A
152, 38



\bibitem[]{}

Taylor, G. B., Peck, A. B., 2003, ApJ 597, 157



\bibitem[]{}

Vermeulen, R. C., Pihlstr\"om, Y. M., Tscharger, W., et al., 2003,
A\&A 404, 861



\bibitem[]{}

Wright, A. and Otrupcek, R., Parkes Catalogue, 1990, Australia
Telescope National Facility



\bibitem[]{}

Xiang, L., Stanghellini, C., Dallacasa, D., Haiyan, Z., 2002, A\&A
385, 768



\bibitem[]{}

Xiang, L., Dallacasa, D., Cassaro, P., Jiang, D., Reynolds, C.,
2005, A\&A 434, 123



\bibitem[]{}

Xu, W., Readhead, A. C. S., Pearson, T. J., Polatidis, A. G.,
Wilkinson, P. N., 1995, ApJS 99, 297



\bibitem[]{}

Yang, J., Liu, X., 2005, ChJAA, Vol.5, Suppl., 224



\end{thebibliography}
\end{document}